\documentclass[fleqn,usenatbib]{mnras}

\usepackage{newtxtext,newtxmath}

\usepackage[T1]{fontenc}

\DeclareRobustCommand{\VAN}[3]{#2}
\let\VANthebibliography\thebibliography
\def\thebibliography{\DeclareRobustCommand{\VAN}[3]{##3}\VANthebibliography}

\usepackage{graphicx}	
\usepackage{amsmath}	
\usepackage{xcolor}
\usepackage{newtxtext,newtxmath}

\newcommand{\wcen}{$\omega$ Cen}
\newcommand{\vphi}{\text{v}_\phi}

\newcommand{\vR}{\text{v}_R}
\newcommand{\vz}{\text{v}_z}
\newcommand{\vx}{\text{v}_x}
\newcommand{\vy}{\text{v}_y}
\newcommand{\vtan}{\text{v}_\text{tan}}
\newcommand{\vmax}{$V_{\text{max}}$}
\newcommand{\rmax}{$r_\text{max}$}

\newcommand{\gaia}{\textit{Gaia}}

\defcitealias{2021arXiv210209568V}{V\&B21}

\title[Dark \& luminous mass components of Omega Centauri]{Dark and luminous mass components of Omega Centauri from stellar kinematics}

\author[Addy J. Evans et al.]{
Addy J. Evans,$^{1}$\thanks{E-mail: addyevans@tamu.edu}
Louis E. Strigari,$^{1}$ Paul Zivick$^{1}$
\\
$^{1}$Mitchell Institute for Fundamental Physics and Astronomy, Department of Physics and Astronomy, Texas A\&M University, College Station, TX\\
}

\date{Accepted XXX. Received YYY; in original form ZZZ}

\pubyear{2021}

\begin{document}
\label{firstpage}
\pagerange{\pageref{firstpage}--\pageref{lastpage}}
\maketitle

\begin{abstract}
We combine proper motion data from \gaia\ EDR3 and HST with line-of-sight velocity data to study the stellar kinematics of the~\wcen~globular cluster. Using a steady-state, axisymmetric dynamical model, we measure the distribution of both the dark and luminous mass components. Assuming both Gaussian and NFW mass profiles, depending on the dataset, we measure an integrated mass of $\lesssim 10^6$ M$_\odot$ within the~\wcen~half-light radius for a dark component that is distinct from the luminous stellar component. For the HST and radial velocity data, models with a non-luminous mass component are strongly statistically preferred relative to a stellar mass-only model with a constant mass-to-light ratio. While a compact core of stellar remnants may account for a dynamical mass up to $\sim 5 \times 10^5$ M$_\odot$, they likely cannot explain the higher end of the range. This leaves open the possibility that this non-luminous dynamical mass component is comprised of non-baryonic dark matter. In comparison to the dark matter distributions around dwarf spheroidal galaxies, the~\wcen~dark mass component is much more centrally concentrated. 
Interpreting the non-luminous mass distribution as particle dark matter, we use these results to obtain the J-factor, which sets the sensitivity to the annihilation cross section. For the datasets considered, the range of median J-factors is $\sim 10^{22} - 10^{24}$ GeV$^2$ cm$^{-5}$, which is larger than that obtained for any dwarf spheroidal galaxy. 
\end{abstract}




\section{Introduction}
\par Omega Centauri (\wcen) is one of the most luminous and well-studied globular clusters (GC) of the Milky Way. It is located at a distance of $\sim$ 5 kpc, is very compact with a half-light radius of $\sim 6$ pc, and has a luminosity of $\sim 2 \times 10^6$ L$_\odot$. There are multiple stellar populations associated with it~\citep{2018ApJ...853...86B,2020ApJ...898..147C}, indicating that it has had a complex star formation history. Indeed, it is possible that~\wcen~is not a GC in the traditionally-defined sense, but rather is the remnant core of a galaxy that has been tidally disrupted~\citep{Wirth:2020dmb}. Additional evidence for this origin is the recently-detected long stellar streams associated with it~\citep{2019NatAs.tmp..258I}, which would be a smoking gun for tidal disruption. 

\par~The dynamical state of~\wcen~has long been a subject of significant interest~\citep{1997AJ....114.1074M}. Utilizing independent dynamical methods, several recent studies of~\wcen~find evidence for an extended mass component that is distinct from that associated with the luminous stellar populations.~\citet{2016A&A...590A..16D} consider a class of non-truncated, radially-anisotropic models, and show that a two-component model fits the data, with a ratio of the ``dark" to luminous mass components of $\sim 1:3$. However, they do not find evidence of mass segregation between the two components.
~\citet{2019MNRAS.482.4713Z} and~\citet{2019MNRAS.488.5340B} also find evidence of a second mass component in~\wcen, and interpret this as evidence for a central cluster of black holes with mass $\sim 5\%$ of the stellar mass. Though not an extended component,~\citet{Noyola2008} find evidence for an intermediate mass black hole (IMBH) with a mass of $\sim 4 \times 10^4$ M$_\odot$ using a spherical and isotropic dynamical model. The subsequent study of~\citet{2010ApJ...710.1063V} sets an upper limit on the mass of an IMBH of $\sim 2 \times 10^4 M_\odot$. 
In addition, recent generalized jeans-based methods~\citep{2010ApJ...710.1063V,2013MNRAS.429.1887D,2013MNRAS.436.2598W} all favor stellar mass-to-light ratios greater than $\sim 2$. 

\par While the dynamics of~\wcen~may be consistent with a population of compact objects, an alternative interpretation is this dark mass distribution is associated with non-baryonic dark matter. This would be natural given the evidence that~\wcen~is associated with the remnant core of a dwarf galaxy. This interpretation has been examined recently by~\citet{Brown:2019whs}, under the assumption that the~\wcen~mass distribution is spherically symmetric. For a Navarro-Frenk-White (NFW) dark matter density profile, these authors find the dynamics are consistent with an integrated mass of $\sim 5 \times 10^5$ M$_\odot$ of dark matter within the half-light radius, amounting to $\lesssim 50\%$ of the stellar mass component. If dark matter does account for this component of the mass distribution, it would be among the most concentrated dark matter distributions measured in any galaxy. 

\par The dynamical evidence for a non-baryonic mass component in~\wcen~has important implications for the interpretation of observations at multiple wavelengths. Among the most intriguing observations of~\wcen~is that from the gamma-ray telescope Fermi-LAT, which has established the existence of an associated gamma-ray point source~\citep{Fermi-LAT:2019yla}. 
This association makes~\wcen~one of the over two dozen GCs that has gamma-ray emission associated with it~\citep{2010A&A...524A..75A}.
Millisecond pulsars (MSPs), known gamma-ray emitters in the Fermi-LAT energy range, have recently been identified in~\wcen~\citep{Dai:2019hkk}, though it is unclear if MSPs can account for the entirety of the gamma-ray emission. Aside from MSPs, it is possible that a second source of the gamma-ray emission in~\wcen~is due to particle dark matter annihilation into Standard Model particles. This possibility has been recently explored by several authors~\citep{Reynoso-Cordova:2019biv,Brown:2019whs}, who find that the gamma-ray spectrum may be fit by a dark matter particle with mass $\sim 30$ GeV and annihilation cross section $\sigma v \simeq 10^{-28}$ cm$^3$ s$^{-1}$. If the source of gamma-rays is indeed from dark matter annihilation, precisely characterizing the mass distribution is crucial in order to obtain the best fits on the particle mass and cross section, and to predict signals from dark matter annihilation across a wide range of wavelengths~\citep{Dutta:2020lqc,Wang:2021hfb}. 

\par In this paper, we model both the dark and luminous mass components of~\wcen. We develop an axisymmetric model based on the CJAM code~\citep{2013MNRAS.436.2598W}, and consider the most updated line-of-sight and tangential velocity dispersion measurements, including the new~\gaia~Early Data Release 3 (EDR3). Though there are some discrepancies in the results derived from the different datasets, we show that all the datasets favor a dynamical mass distribution that includes a dark component of order $\sim 10^4 - 10^6$ M$_\odot$.
Though it is detected at a statistically-significant level, this component is still subdominant relative to the dynamical influence of the luminous stellar component. 

\par This paper is organized as follows. In Section~\ref{sec:methods}, we discuss our axi-symmetric, steady-state model for~\wcen. In section~\ref{sec:profiles} we introduce the models for the dark mass component. In section~\ref{sec:data} we review the datasets that are used in our analysis, and in section~\ref{sec:multinest} we discuss the nested sampling routine utilized to analyze the data. In section~\ref{sec:results}~and section~\ref{sec:discussion}~we discuss the results of our analysis and their implications.

\section{Theoretical methods}
\label{sec:methods}
We use CJAM to model the photometry and the stellar kinematics of~\wcen. CJAM~\citep{2013MNRAS.436.2598W} is an extension of Jeans Anisotropic Modeling (JAM) introduced by~\citet{2008MNRAS.390...71C}, which utilizes the axisymmetric Jeans equations to calculate first and second velocity moments. Assuming that the system is axisymmetric and in steady-state ($\frac{\partial}{\partial \phi} = 0 \text{ and } \frac{\partial}{\partial t} = 0$), using a cylindrical polar coordinate system the second moment Jeans equations can be written as
\begin{equation}
    \frac{\nu \left (\overline{\vR^2} - \overline{\vphi^2} \right )}{R} + \frac{\partial \left ( \nu \overline{\vR^2} \right )}{\partial R} + \frac{\partial \left ( \nu \overline{\vR \vz} \right )}{\partial z} = -\nu \frac{\partial \Phi}{\partial R} 
    \label{eq:jeans}
\end{equation}
\begin{equation}
    \frac{\nu \left(\overline{\vR \vz} \right)}{R} + 
    \frac{\partial \left ( \nu \overline{\vR \vz} \right )}{\partial R} +
    \frac{\partial \left ( \nu \overline{\vz^2} \right )}{\partial z} =
    -\nu \frac{\partial \Phi}{\partial z}
    \label{eq:jeans2}
\end{equation}
Here $\nu$ is the 3D (de-projected) luminosity density of the system, ($\vR$, $\vphi$,$\vz$) are velocities in their respective coordinate directions, and $\Phi(R,z)$ is the gravitational potential of the system. The overline denotes an average of the given quantity over the stellar distribution function, with first moments taking the form $\overline{{\rm v}}_\imath$, and second moments taking the form $\overline{{\rm v}^2_{\imath \jmath}}$, where $\imath, \jmath = R, \phi, z$. For our axisymmetric assumptions, we have $\overline{{\rm v}_R} = \overline{{\rm v}_z} = 0$, and similarly for the following mixed second moments, $\overline{{\rm v}_{R} {\rm v}_\phi} = \overline{{\rm v}_\phi {\rm v}_z} = 0$. As is standard, we define the velocity anistropy of the system as
\begin{equation}
    \beta = 1 -  \frac{\overline{{\rm v}_z^2}}{\overline{{\rm v}_R^2}}. 
\end{equation}

\par CJAM models the luminosity density of the stellar distribution as a sum of $N$ Gaussians terms  
\begin{equation}
    \nu(R,z) = \sum_{k=1}^N \frac{L_k}{(2\pi \sigma_k^2)^{3/2}q_k } \exp \left [-\frac{1}{2\sigma_k^2}\left(R^2 + \frac{z^2}{q_k^2} \right) \right], 
    \label{eq:luminositydensity}
\end{equation}
where $q_k$ gives the intrinsic flattening of the Gaussian component. The quantity $\sigma_k$ may be identified as the characteristic radius of a Gaussian component. Defining a coordinate system with $(x^\prime,y^\prime)$ in the plane of the sky, the positive $z^\prime$ direction away from the observer, and projecting the luminosity onto the plane of the sky gives the surface brightness as a sum of the Gaussian components,
\begin{equation}
    \Sigma(x^\prime,y^\prime) = \sum_{k=1}^N \frac{L_k}{2\pi \sigma_k^2 q^{'}_k } \exp \left [-\frac{1}{2\sigma_k^2}\left(x^{'2} + \frac{y^{'2}}{q^{'2}_k} \right) \right]  
    \label{eq:surfacedensity}
\end{equation}
where $q_k^\prime$ now gives the projected flattening of each Gaussian component, given by its relationship to $q_k$ and the inclination angle $i$,
\begin{equation}
    q_k = \frac{\sqrt{q_k'^2 - \cos^2 i}}{\sin i}
    \label{eq:flatteningratio}
\end{equation}
\par~\citet{2013MNRAS.429.1887D} have fit the surface brightness profile of~\wcen~to the Multi-Gaussian Expansion (MGE) model of Equation~\ref{eq:surfacedensity}, with the results reproduced in Table 1. These fits imply that~\wcen~ is very close to spherical, as most of the Gaussian components have $q_k^\prime \simeq (0.9-1.0)$. We note that this surface brightness profile is consistent, both in terms of the shape and the normalization, with the parameterized function of~\citet{2010ApJ...710.1063V} and with the data compiled by these authors. 

\par From the luminosity density, we can proceed to define the stellar mass density from the sum of the Gaussian components,
\begin{equation}
    \rho_\star (R,z) = \sum_{j=1}^{M_j} \frac{M_j}{(2\pi\sigma_j^2)^{3/2}q_j} \exp \left[-\frac{1}{2\sigma_j^2} \left( R^2 + \frac{z^2}{q_j^2}\right)\right].
    \label{eq:massdensity}
\end{equation}
The contribution to the total mass density of the system from the stellar component can then be determined by defining the stellar mass-to-light ratio, $\Upsilon_{*}$, which gives the stellar mass $M_j$ associated with each Gaussian component, 
\begin{equation}
    M_j = 2\pi L_j \sigma_j^{2}q_j^{'} \Upsilon_{*}
\end{equation}
where $2\pi L_j \sigma_j^2 q_j^{'}$ gives the total luminosity of each Gaussian component and, in this case, $j = k$. Note that with this definition, the mass-to-light ratio is the same for each of the Gaussian components. 

\par For the analysis in this paper, we will be interested in the spherically-averaged stellar mass profile within a radius $r = \sqrt{R^2 + z^2}$. To determine this we first define the spherically-averaged stellar density profile as 
\begin{equation}
    \overline{\rho}_\star(r) = \frac{1}{r} \int_0^{r} \rho_\star \left(\sqrt{r^2-z^2},z \right) dz. 
    \label{eq:sphericallyaveragedmass}
\end{equation}
From this we obtain the spherically-averaged stellar mass profile within a radius $r$ as 
\begin{equation}
    M_\star(r) = \int_0^{r} 4 \pi r^{\prime 2} \overline{\rho}_\star(r^\prime) dr^\prime. 
\end{equation}

\par In addition to contributions from the luminous component, Equations~\ref{eq:jeans}~and~\ref{eq:jeans2}~accommodate contributions to the potential $\Phi$ from non-luminous components. In the CJAM formalism, these non-luminous components may also be parameterized as a sum of Gaussian components. As described below, we include a dark mass component which is parameterzied as a sum of Gaussians, which includes well-motivated models for the density profile of dark matter halos. 

\par To compare to the measured proper motion and line-of-sight velocity dispersions, we must transform the first and second velocity moments to the ($x^\prime, y^\prime, z^\prime$) coordinate system. We further can define the inclination of the system, $i$, such that $i = 0^{\circ}$ for face-on systems and $i=90^{\circ}$ for edge-on systems. Following these transformations, we can derive first and second velocity moments for the $x^\prime$,$y^\prime$, and $z^\prime$ components~\citep{1994MNRAS.271..202E,2013MNRAS.436.2598W}: 
\begin{eqnarray}
    \label{eq:vmoments} 
    \overline{\vx{}_{\prime}} &=& \overline{\vR} \cos{\phi} - \overline{\vphi} \sin{\phi} \\
    \overline{\vy{}_{\prime}} &=& - \left( \overline{\vR} \sin{\phi} + \overline{\vphi} \cos{\phi} \right )\cos{i} + \overline{\vz}\sin{i} \nonumber \\ 
    \overline{\vz{}_{\prime}} &=& \left( \overline{\vR} \sin{\phi} + \overline{\vphi}\cos{\phi}\right)\sin{i} + \overline{\vz} \cos{i} \nonumber \\
    \overline{v^2_{x^{\prime}}} &=& \overline{v_R^2}\cos^2\phi  + \overline{v_\phi^2}\sin^2 \phi \nonumber \\
    \overline{v^2_{y^{\prime}}} &=& \left(\overline{v_R^2}\sin^2\phi + \overline{v_\phi^2}\cos^2 \phi\right)\cos^2 i + \overline{v_z^2}\sin^2 i \nonumber \\
    \overline{v^2_{z^{\prime}}} &=& \left(\overline{v_R^2}\sin^2\phi + \overline{v_\phi^2}\cos^2 \phi\right)\sin^2 i + \overline{v_z^2}\cos^2 i \nonumber 
\end{eqnarray}
The above equations assume that the velocity ellipsoid is aligned with the cylindrical polar coordinate system, so that $\overline{{\rm v}_R {\rm v}_z} = 0$.
We choose this alignment both for simplicity and to follow the formalism presented in in~\citet{2008MNRAS.390...71C}~and~\citet{2013MNRAS.436.2598W}. However, we note that other alignments are possible, e.g. the spherical alignment discsussed in~\citet{2020MNRAS.494.4819C}.

\par The final step in the comparison to observations involves integrating the first and second moments along the $z^\prime$ axis. For the first and second velocity moments in Equation~\ref{eq:vmoments}, respectively, these integrated moments are
\begin{eqnarray}
\mu_{\imath} = \frac{1}{\Sigma} \int d z^\prime \rho_\star \overline{{\rm v}_\imath} \\
\overline{\mu_{\imath}^2} = \frac{1}{\Sigma} \int d z^\prime \rho_\star \overline{{\rm v}_\imath^2}, 
\end{eqnarray}
where here the subscript refers to the coordinates $\imath = x^\prime, y^\prime, z^\prime$. 

\par In addition to the first and second moments above, mixed second moments may be considered in a dynamical analysis. However, as we discuss below, since we analyze data for the most part in the form of velocity dispersions and not individual velocities, we will not include mixed second moments in our analysis.

 \par The second moments above do not explicitly split into contributions from streaming motions and random motions. Performing this decomposition requires a model for the rotation of the system.~\citet{1980PASJ...32...41S} and ~\citet{2008MNRAS.390...71C} discuss a rotation model, which is implemented in CJAM, relating the streaming motion in the $\phi$ direction to the second moments in the $R$ and $z$ directions,
 \begin{equation}
     [{\cal L}\overline{\vphi}]_k = \kappa_k \left(\left[{\cal L}\overline{\vphi^2}\right]_k - \left[{\cal L}\overline{v_R^2}\right]_k\right)^{1/2}. 
     \label{eq:kappaeqn}
 \end{equation}
 Here the parameter $\kappa_k$ is defined such that $\kappa_k = 0$ for no streaming motion, and $|\kappa_k| = 1$ for one with a circular velocity ellipsoid. The rotation parameter is largely dependent upon the tangential velocities of the system. Given our assumption of the position angle of the system as defined in~\citet{2013MNRAS.436.2598W}, a positive tangential velocity value yields a negative $\kappa$ value and vice versa.

\par
From the first and second velocity moments we can define the velocity dispersion in a given coordinate direction,
\begin{equation}
    \sigma_{\imath}^2 = \overline{v_{\imath}^2} - \overline{v_\imath}^2. 
\end{equation}
Three independent components of the velocity dispersion are then obtained: two proper motion components in the plane of the sky, and one along the line of sight. These are functions of the projected coordinates $x^\prime, y^\prime$, and are used to compare to the corresponding quantities constructed from the data. 

\par As described below, the velocity dispersion at a given position is approximated from the data by binning stars in the projected radial coordinate $R$.  To compare to all the datasets, we will be interested in the azimuthally-averaged line-of-sight and proper motion velocity dispersion profiles. From the second moment equations above, we determine the azimuthally-averaged profile for a particular component as
 \begin{equation}
     \langle \sigma_\imath^2 (R) \rangle = \frac{1}{2\pi}\int_0^{2\pi} \sigma_\imath^2  (R,\eta)\text{d}\eta,
     \label{eq:sigmaaz}
 \end{equation}
 where $\eta$ is an angle in the plane of the sky defined from the positive $x^\prime$-axis. 
Here $\sigma_\imath (R,\eta)$ are the dispersions calculated for each value of $(R,\eta)$ from CJAM, and $\langle {\sigma}_\imath^2 (R) \rangle$ is the azimuthally averaged velocity dispersion. For data with line-of-sight velocity measurements, we take the line-of-sight velocity dispersion to be the RMS value of the velocity dispersion along the line of sight: $ \sigma_{LOS} = \sqrt{\langle \sigma^2_{z'} \rangle }$. For the datasets with proper motions, we will need to construct the velocity dispersion averaged over directions in the plane of the sky. From the second moments defined above, this average 1D velocity dispersion may be written as $\sigma_{1D} = [ (\langle \sigma_{x^\prime}^2 \rangle + \langle \sigma_{y^\prime}^2 \rangle)/2 ]^{1/2}$. 

\par As we discuss below, the \gaia\ EDR3 data is the only dataset we utilize with absolute measurements for proper motions of the stars. We use this dataset to constrain the intrinsic rotation of~\wcen~using Equation~\ref{eq:kappaeqn}. For this analysis, we bin the velocities in positions $R$, and as with the dispersions calculate the azimuthally-averaged velocity in each radial bin as
\begin{equation}
     \langle {\rm v}_\imath (R) \rangle = \frac{1}{2\pi}\int_0^{2\pi} {\rm v}_\imath  (R,\eta)\text{d}\eta,
     \label{eq:vaz}
 \end{equation}
 We implement Equation~\ref{eq:sigmaaz} and~\ref{eq:vaz} in our likelihood analysis described in Section~\ref{sec:multinest}. 
  
 \section{Density profiles} 
 \label{sec:profiles}
\par To properly add a dark mass (DM) component to the mass density MGE, we must choose a density profile, and then describe this profile as a summation of Gaussian components. An often-used and well-motivated dark matter distribution in galaxies is the NFW~\citep{1996ApJ...462..563N} profile, which describes a cuspy density distribution. For an NFW profile, the density is given by:
\begin{equation}
\rho_{\text{NFW}}(r) = \frac{\rho_s}{\frac{r}{r_s}\left(1+\frac{r}{r_s}\right)^2}
\label{eq:NFW} 
\end{equation}
and the gravitational potential is defined as
\begin{equation}
    \frac{\Phi_{\text{NFW}}(r)}{\Phi_s} =  1 - \frac{\ln \left(1+\frac{r}{r_s}\right)}{\frac{r}{r_s}}.
\end{equation}
Here $\Phi_s = 4\pi G \rho_s r_s^2$ and $\rho_s$ and $r_s$ are defined as the characteristic density and characteristic radius of the dark component. For our purposes, it is more convenient to describe the density in terms of the maximum circular velocity, $V_{\text{max}}$, and the radius at which this is attained, $r_{\text{max}}$. 
We can use the following relations to rewrite equation~\ref{eq:NFW} in terms of $V_{\text{max}}$ and  $r_{\text{max}}$:
\begin{equation}
    r_{\text{max}} = 2.16 r_s  \hspace{1cm} V_{\text{max}} = 0.465\sqrt{\Phi_s}
\end{equation}

\par The NFW model is restricted to describe cuspy density profiles; we have checked that the assumption of a centrally-cored profile, such as a Burkert model~\citep{1995ApJ...447L..25B}, does not affect our conclusions. As we found no difference in our results between an NFW or Burkert profile, we stick with the NFW model for our fiducial analysis.

\par
\begin{figure} 
    \includegraphics[width=\columnwidth]{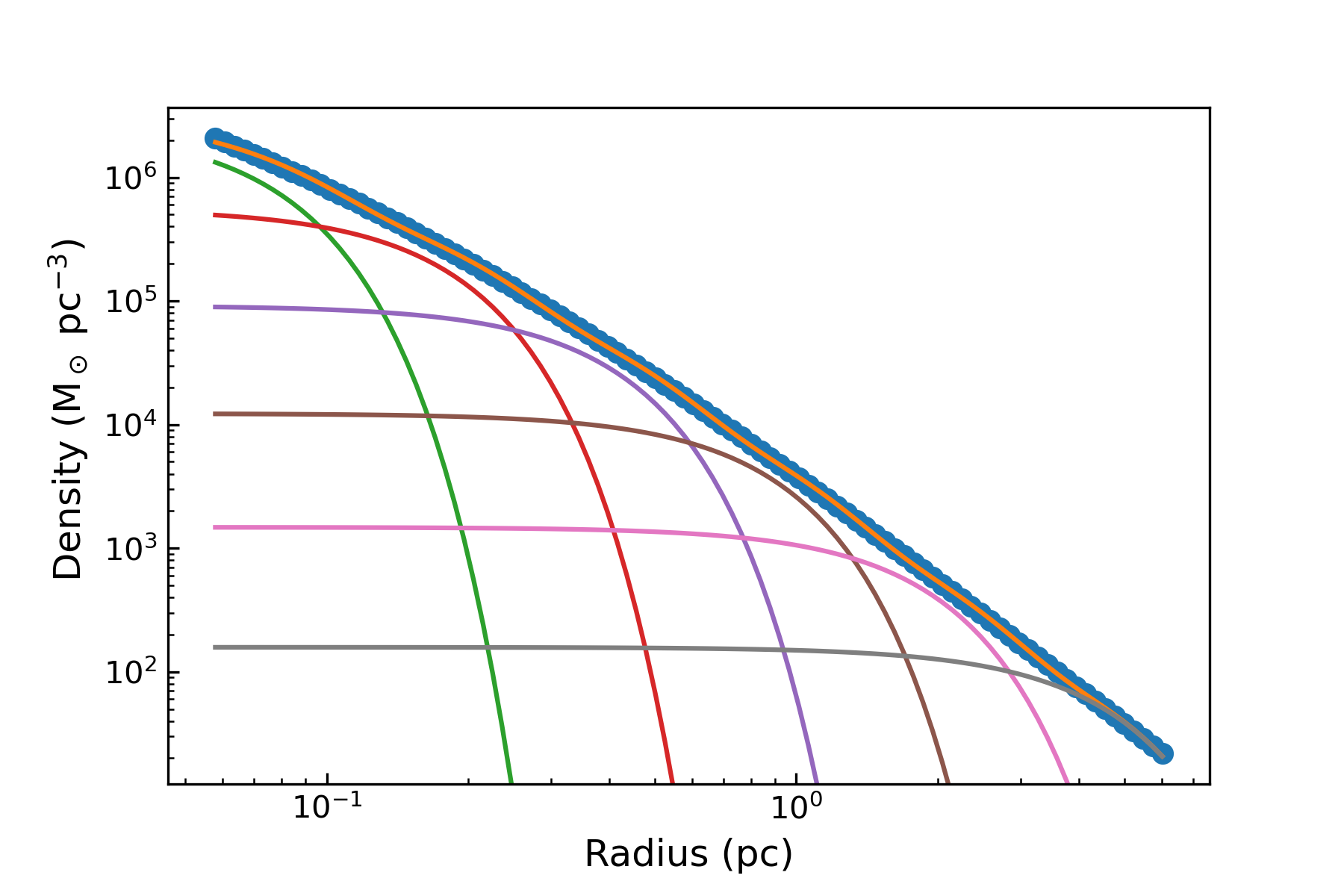}
    \caption{Gaussian fitting to an NFW profile using MGEFit for typical values of~\vmax~and~\rmax~from our results. The profile is fit using the maximum and minimum radial distances from each dataset, assuming a distance to the cluster of 5240 pc. Here, the HST dataset range is shown for an NFW profile with a value of~\vmax~$ = 20$ km/s and~\rmax~$ = 0.32$ pc.}
    \label{fig:mgefit}
\end{figure}

 In order to utilize the NFW density profiles with CJAM described above, we must fit the density profile to a sum of Gaussian components. To fit Gaussian components to our NFW density profiles, we use the MGE fitting method described in~\citet{2002MNRAS.333..400C}. The parameters which describe these Gaussian curves, namely the density and the characteristic width of the curve, are used to build upon the MGE description of the mass density profile of~\wcen. The MGE fitting method allows for the density profile to be fit by an arbitrary number of Gaussians. For our analysis, we fix to six Gaussian components, as this number provides an optimal balance between speed of calculation and accuracy of the fit. The typical residuals we find in our fits are $\lesssim 10\%$. Figure \ref{fig:mgefit} shows an example of using MGE fitting on an NFW profile over the radial range of the HST data (described below) with a~\vmax~of 20 km/s and an~\rmax~of 0.32 pc, which as we show later are typical values for these parameters from our analysis. 
\par The NFW profile has an infinite mass, and therefore ultimately cannot fully describe a physical density profile. Further, the $\sim r^{-3}$ fall-off in the outer region of the halo does not account for tidal stripping, which has almost certainly occurred for the case of~\wcen. To account for these effects, we compare the NFW distribution to a Gaussian distribution. This distribution is described by the addition of a single Gaussian component, given by~\citet{2013MNRAS.436.2598W}:
\begin{equation}
\rho_{\text{Gaussian}}(r) = \frac{M}{(2\pi\sigma^2)^{3/2}q}\exp{\left[-\frac{1}{2\sigma^2}\left(R^2 + \frac{z^2}{q^2}\right)\right]}
\end{equation}
Here, as for the case of the stellar distribution, $M$ is the total mass of the dark component, $\sigma$ is the Gaussian width, and $q$ is the projected flattening ratio which we take to be $1$ (the spherical distribution case). We compare the results of the NFW case to the Gaussian case in section~\ref{sec:results}. In the limit of small $\sigma$, this distribution may be used to model a central IMBH. Moreover, we choose this density profile for its simplicity to implement within CJAM and its capability to test an extended dark component as well as the case of an IMBH.
 
\begin{table}
	\centering
	\caption{The multi-Gaussian expansion of $\omega$ Cen's surface brightness profile, derived by~\citet{2013MNRAS.429.1887D} and used by~\citet{2013MNRAS.436.2598W}. Here, $k$ is the Gaussian number, $L_k$ is the surface brightness, $q_k'$ is the flattening ratio, and $\kappa_k^{\alpha}$ is the rotation parameter. The $\kappa$ values we use for our analysis are opposite in sign as compared to the values determined in~\citet{2013MNRAS.429.1887D} due to the position angle used in~\citet{2013MNRAS.436.2598W}.}
	\label{tab:surfaceBrightnessMGE}
	\begin{tabular}{lcccr} 
		\hline
		$k$ & $L_k$ & $\sigma_k$ & $q_k'$ & $\kappa_k^{\alpha}$\\
		& (L$_\odot$pc$^{-2}$) & (arcmin) & & \\
		\hline 
		1 & 1290.195 & 0.475570 & 1.0000000 & 0.0\\
		2 & 4662.587 & 1.931431 & 0.9991714 & 0.0\\
		3 & 2637.784 & 2.513385 & 0.7799464 & -0.4\\
		4 & 759.8591 & 3.536726 & 0.7241260 & -1.1\\
		5 & 976.0853 & 5.403728 & 0.8556435 & -0.6\\
		6 & 195.4156 & 8.983056 & 0.9392021 & 0.0\\
		7 & 38.40327 & 13.93625 & 0.9555874 & 0.0\\
		8 & 8.387379 & 20.98209 & 1.0000000 & 0.0\\
		
		\hline
	\label{table:1}
	\end{tabular}
\end{table}

\section{Data}
\label{sec:data} 
In this section we describe the datasets that we use in our analysis. 
We use data from the publicly available Baumgardt Globular Cluster Database \footnote{\href{https://people.smp.uq.edu.au/HolgerBaumgardt/globular/}{Baumgardt Globular Cluster Database}} and the catalog from \gaia\ EDR3 in conjunction with the globular cluster membership analysis in~\citet{2021arXiv210209568V} (hereafter \citetalias{2021arXiv210209568V}).
The Baumgardt Globular Cluster Database is a compilation of up-to-date proper motions (PMs) and line-of-sight velocities and dispersions for 159 Milky Way GCs. 
From this, we use a combination of dispersion measurements from Hubble Space Telescope (HST) PMs and from ground-based radial velocities (RV) to describe the stellar kinematics of $\omega$ Cen. 
Additionally, using the EDR3 catalog and selecting individual stars on a membership cut from \citetalias{2021arXiv210209568V}, we directly calculate the dispersions and rotational PMs of $\omega$ Cen.

As a general point we choose to fit to the datasets used in this analysis separately, and not combine them into a single analysis. Combining the data sets would require the introducing of hyperparameters in our Bayesian analysis, which is beyond the scope of this present study. In addition, the different data sets may introduce different sets of systematics that are difficult to quantify in a combined analysis. As described at the end of Section~\ref{sec:methods}, we also choose to use binned velocity dispersion data in our analysis for all datasets, for both computational convenience, and because this is the format that the HST data is presented in.
\par 
It is worth noting that while the \gaia\ and HST data both offer us information about motions in the plane of the sky, they cover vastly different spatial scales. 
Due to the stellar crowding at the center of globular clusters, \gaia\ does not reliably provide high quality astrometric measurements in those regimes. 
As such, for the case of ~\wcen, after applying the astrometric quality cuts recommended in \citet{Fabricius2021}, the innermost remaining star is located roughly 3.6 arcminutes from the center or about 5.5 parsecs.
On the other hand, HST is able to resolve individual stars within $\sim 0.05$ parsecs of the center. 
\par As one final note, we assume a distance of 5.24 kpc for all datasets to match the distance assumed in~\citet{2019MNRAS.482.5138B} along with the corresponding center (($\alpha$, $\delta$) (J2000) = ($201.696838^\circ$,$-47.479584^\circ$)) for our tangential velocity calculations.

\subsection{\gaia\ EDR3}
\par \gaia\ EDR3 offers an unparalleled dataset of internal PMs in~\wcen. The new catalog extends the observations used for analysis from 22 months in Data Release 2 to 34 months in EDR3. The longer baseline, increased sampling, and the opportunity to improve the astrometric reference frame of ~\gaia~has resulted in improved statistical uncertainties for individual stars (approximately a factor of $\sim2$ for PM uncertainties, see~\citealt{Lindegren2021a}) in addition to smaller systematic uncertainties (with a floor of about $\sim0.026$ mas yr$^{-1}$ as found in  \citetalias{2021arXiv210209568V}).
 
\par As our focus in this work is on the dynamic modeling of~\wcen, we defer to the membership analysis presented in \citetalias{2021arXiv210209568V} for selecting stars out of the EDR3 catalog. 
Using their final published catalog that contains membership probabilities for individual stars, we select only those stars which have a membership probability greater than $99\%$. 
For full details on their methodology, we refer the reader to \citetalias{2021arXiv210209568V}. 
However, as a brief summary, they make a series of astrometric quality cuts, based on the recommendations from \citet{Fabricius2021}, to select astrometrically well-behaved stars. 
With this cleaned catalog, they use a multi-Gaussian mixture model to describe the combination of foreground Milky Way stars and the target cluster, using the Markov Chain Monte Carlo code EMCEE \citep{Foreman-Mackey13} to explore the parameter space and calculate the membership probabilities for individual stars using the converged MCMC runs. 
After applying our high-probability cut, we are left with a final sample of nearly 50,000 stars that are consistent with belonging to ~\wcen.

\par Our first step in working with this trimmed sample is to measure the 2 dimensionally-averaged (hereafter just 2D) PM dispersion as a function of cluster radius in order to create a set of dispersions similar to the HST 2D dispersion measurements. 
To do this, we first calculate the radial distance from each star to the center of the cluster (assuming the center discussed above for ~\wcen) using the Cartesian transformations laid out in \cite{2018A&A...616A..12G}. 
Because we are simply looking to calculate the 2D PM dispersions, we leave the PMs for each star in the original ($\mu_{\alpha}$, $\mu_{\delta}$) values in order to avoid introducing any potential systematics into the dispersion measurement.

\par To begin the measurement, we first set the number of stars per bin such that the statistical uncertainty on the dispersion and average PM of the bin is equivalent to the systematic unceratinty found in \citetalias{2021arXiv210209568V} in order to fully leverage the information in our sample. 
This results in $\sim350$ stars per bin for our analysis, beginning with the innermost star and counting outwards (e.g., the innermost 350 stars are the first bin, the next 350 are the second bin and so on). For each bin, the PMs are modeled as a multi-variate Gaussian:
 \begin{equation}
 \label{eq:pmlike}
     \mathcal{L}_{\mathrm{pm}} = (2\pi)^{-1/2} (\det\Sigma)^{-1/2} \exp{\displaystyle[ -\frac{1}{2} (\mu - \overline{\mu})^{T} \Sigma^{-1} (\mu - \overline{\mu})}]
 \end{equation}
where $\mu = (\mu_{\alpha}\cos{\delta}, \mu_{\delta})$ is the data vector (\texttt{pmra} and \texttt{pmdec}, respectively, in the EDR3 catalog) and $\overline{\mu} = (\overline{\mu_{\alpha}\cos{\delta}}, \overline{\mu_{\delta}})$ is the systemic PM of the stars in the bin. 
The covariance matrix $C$ includes the correlation between the proper motion uncertainties ($\epsilon$, \texttt{pmra\_error} and \texttt{pmdec\_error} in EDR3) along with a term for the intrinsic PM dispersion of the binned stars ($\sigma$):
\begin{equation}
\label{eq:pmcovar}
    C = \begin{bmatrix}
    \epsilon^2_{\mu_{\alpha}\cos{\delta}} + 
    \sigma^2_{\mu_{\alpha}\cos{\delta}} & 
    \rho \epsilon_{\mu_{\alpha}\cos{\delta}} \epsilon_{\mu_{\delta}} \\ 
    \rho \epsilon_{{\alpha}\cos{\delta}} \epsilon_{\mu_{\delta}} &
    \epsilon^2_{\mu_{\delta}} +
    \sigma^2_{\mu_{\delta}}
    \end{bmatrix}
\end{equation}
where $\rho$ is the correlation between the PM measurements, provided in the EDR3 catalog as \texttt{pmra\_pmdec\_corr}.

We then estimate the best-fit values for $\overline{\mu_{\alpha}\cos{\delta}}, \overline{\mu_{\delta}}, \sigma_{\mu_{\alpha}\cos{\delta}},$ and
$\sigma_{\mu_{\delta}}$ using the~\texttt{emcee}~library \citep{Foreman-Mackey13}, which implements the affine-invariant ensemble sampler for Markov Chain Monte Carlo~\citep{Goodman&Weare10} in Python.
While we do not expect the average PM of the cluster to substantially change as a function of radius, fixing the cluster's PM to a single value across all bins could potentially introduce new systematics to the dispersion measurement.
Ultimately, as we are concerned primarily with the intrinsic width of the the PM distribution and not its exact center, we prefer to fit for both the average PM and the intrinsic dispersion.
The resulting best-fit dispersion values in each PM direction are then combined into a 2D-averaged dispersion (as with HST) and used in our modeling (shown as the purple points in Figure \ref{fig:observed_dispersions}). 
As a consistency check, we find our final dispersions in good agreement with the dispersions presented in \citetalias{2021arXiv210209568V}, although slight differences are present likely due to the difference in how the stars were radially binned ($\sim12$ bins in \citetalias{2021arXiv210209568V} compared to our 100+ bins).

As a final note on the dispersions, we recognize the potential for a small complicating effect from the change in how the systemic 3D motion of \wcen~projects into the plane of the sky. 
Because the dispersion is ideally measured in the frame of the studied object, a change in how the systemic motion changes as a function of position from the center of the object, this could impact the dispersion. 
To check the potential of this, we apply the formalism of \citet{vanderMarel2002} to correct for this geometric perspective effect and recalculate the dispersions as a function of radius. 
We find this correction for \wcen\ to have a negligible effect on the resulting dispersions. As such, we do not expect this to affect our analysis, and we defer to the original dispersion calculation using only the \gaia-provided ($\mu_{\alpha}$, $\mu_{\delta}$) in providing constraints on our dynamic models.

As mentioned earlier, the second quantity we wish to calculate using the \gaia\ data is the rotational PMs as a function of radius.
To start, using the Cartesian projection discussed above, we transform the position and PM of each star from ($\alpha$, $\delta$) (J2000), ($\mu_{\alpha}$, $\mu_{\delta}$) into ($X, Y$), ($\mu_X, \mu_Y$).
Now with this projection, ($X, Y$) are centered relative to the given cluster center, but ($\mu_X, \mu_Y$) are not yet relative to the cluster, which we will need to calculate the rotational PM.
To do so, we take the mean PM of \wcen~from Table A1 in \citetalias{2021arXiv210209568V} and transform it into the Cartesian frame as well, giving us ($\overline{\mu_X}$, $\overline{\mu_Y}$).
We then subtract this systemic PM from the Cartesian PM of each star to produce a set of relative PMs.

From here we transform these relative PMs, now that both position and PM are centered on the cluster, into a cylindrical polar coordinate system which each star is described by ($R$, $\phi$), ($\mu_R, \mu_\phi$) (where $R$ was used earlier in binning the stars for the dispersion calculation, and $\mu_R$ and $\mu_\phi$ are the radial and rotational PM components).
Then, as with the 2D dispersions, we radially bin the stars, beginning with the innermost 350, and so on, and calculate the average rotational PM in each bin (shown as the purple points in Figure \ref{fig:modelVtans}).
As discussed above, when attempting to determine relative PMs for an extended body in the plane of the sky, there is a complication due to the changing projection of the systemic motion of the object across the sky.
However, as noted in \citet{2006A&A...445..513V}, this only affects the radial PM component, not the rotational PM component, so as we are only focused on the rotational PM component, we do not apply this correction.

\subsection{HST}
Proper motions of $\omega$ Cen from HST were determined by~\citet{2014ApJ...797..115B} using archival HST images taken with WFC3/UVIS, ACS/WFC, and ACS/HRC. 
Further details on the cameras can be found in the aforementioned paper. 
In particular, data reduction and proper motion determinations were made for $\omega$ Cen using twelve observation epochs, across which stars were cross-identified and their positions eventually transformed onto a common reference frame to obtain final proper motions. 
The velocity dispersions of these measurements were then derived in~\citet{2015ApJ...803...29W} using a maximum likelihood method. 
\subsection{RV}
The radial velocities used in our analysis are a combination of spectra from ESO/VLT and Keck compiled in~\citet{2018MNRAS.478.1520B}. The authors also used published literature of line-of-sight data compiled in~\citet{2017MNRAS.464.2174B}.
The individual stellar radial velocities were then cross-correlated across the datasets to determine mean radial velocities. 
Similarly to the other two datasets, a maximum likelihood calculation was used to determine the final velocity dispersion profile.

\subsection{MUSE}
\par In addition to the RV data described above, radial velocity measurements have been made in the central regions of~\wcen, at $\lesssim 1$ arcmin by~\citet{2018MNRAS.473.5591K} as well as~\citet{2008ApJ...676.1008N}. While there is significant scatter in the measurements between these datasets, we choose to include the data from~\citet{2018MNRAS.473.5591K} (hereafter MUSE) as it is the most up-to-date LOS data available. We choose to keep the MUSE data separate from the other RV measurements as the MUSE data is the only set taken with an integrated field spectrograph. The particular data that we use was taken in October 2016 as part of a larger survey of Milky Way globular clusters. The authors of this survey took 10 pointings of~\wcen, roughly corresponding to 10 arcminutes of coverage. Data on~\wcen~was taken over four epochs for a total observation time of 4.9 hours. Similar to the other data sets discussed, a maximum likelihood approach is utilized to determine velocity dispersions.

\par
\begin{figure}
	\includegraphics[width=\columnwidth]{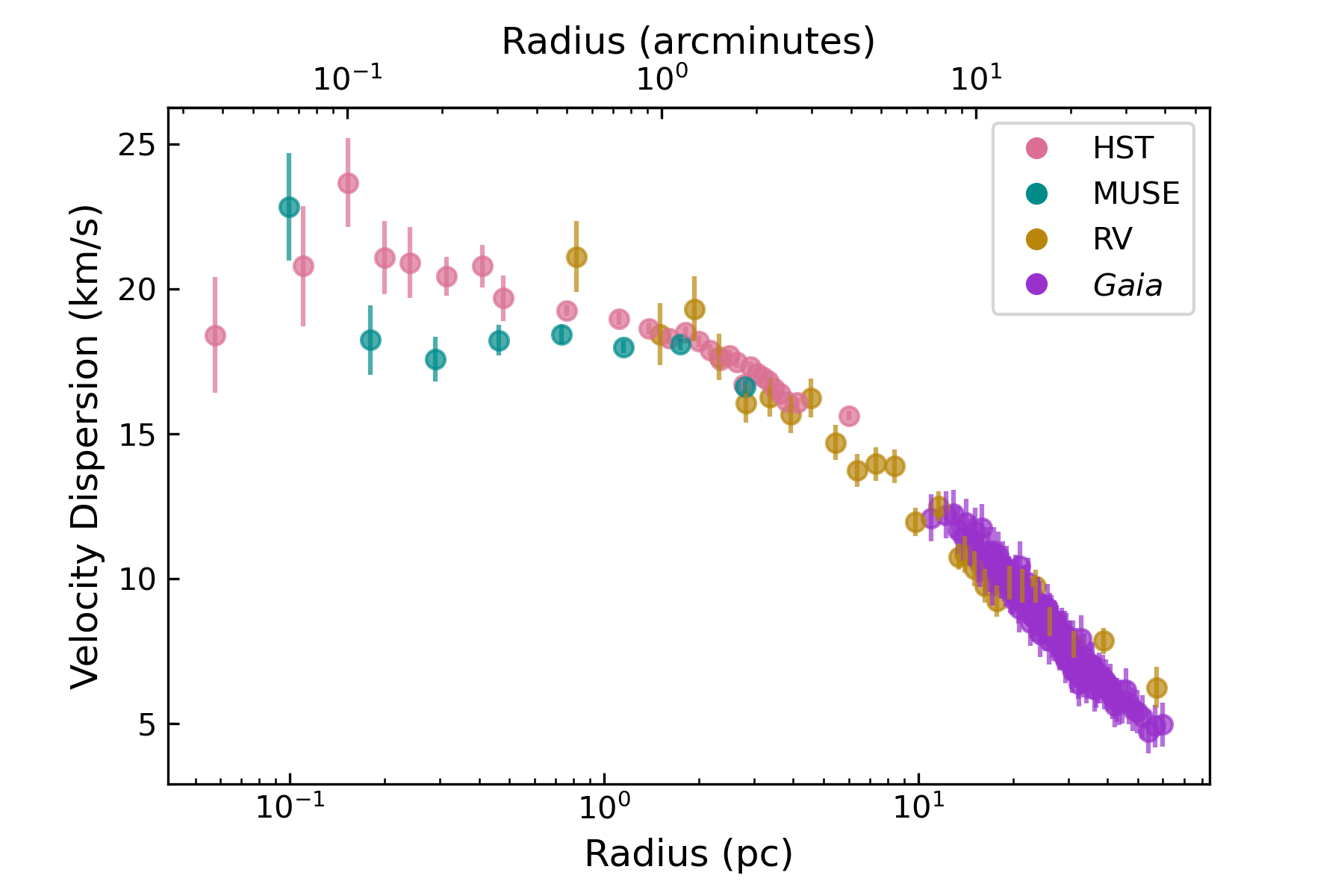}
    \caption{The datasets used in our analysis. Proper motion dispersions from \textit{Gaia} ~\citep{2021arXiv210209568V} (purple), line-of-sight dispersions compiled in~\citet{2018MNRAS.478.1520B} and~\citet{2017MNRAS.464.2174B} (yellow), HST proper motion dispersions derived in~\citet{2015ApJ...803...29W} (pink), and MUSE IFS radial velocity dispersions from~\citet{2018MNRAS.473.5591K}.}
    \label{fig:observed_dispersions}
\end{figure}

 Figure \ref{fig:observed_dispersions} shows the cumulative velocity dispersion profile from all four datasets used in our analysis. The combination of datasets spans from as close to $\sim 0.05$ pc from the center of the cluster out to well beyond the half-light radius at a distance from the center of $\sim 60$ pc. 

\section{Bayesian inference with MultiNest}
\label{sec:multinest}
\par To compare our model to the data, we bin both the velocity dispersions and the tangential velocities in projected radius $R$. We then define the following likelihood function:
\begin{align}
   \mathcal{L(M)}_\alpha = \prod_\imath \frac{1}{\sqrt{2\pi \epsilon_{i,\text{disp}}^2}}\exp{\left[-\frac{(\sigma_{\text{obs}}(R_i) - \sigma_{\text{CJAM}}(R_i))^2}{2\epsilon_{i,\text{disp}}^2}\right]} \nonumber \\ 
   \times \frac{1}{\sqrt{2\pi \epsilon_{i,\vtan}^2}}\exp{\left[-\frac{(\vtan{}_{\text{obs}}(R_i) - \vtan{}_{\text{CJAM}}(R_i))^2}{2\epsilon{}_{i,\vtan}^2}\right]}
   \label{eq:vtanlikelihood}
\end{align}
where $\alpha$ represents one of the four datasets that we utilize. Here $\epsilon{}_i$ is the error associated with the $i$th measurement and $\sigma{}_{\text{obs}}(R_i)$ and $\vtan{}_{\text{obs}}(R_i)$ are the observed dispersion and tangential velocity of this measurement in a bin centered at the $i$th radius. Likewise, $\sigma{}_{\text{CJAM}}(R_i)$ and $\vtan{}_{\text{CJAM}}(R_i)$ are the CJAM azimuthally-averaged dispersion and tangential velocity at the $i$th radius from Equations~\ref{eq:sigmaaz} and~\ref{eq:vaz}.

\par The first term in Equation~\ref{eq:vtanlikelihood} that depends on the binned velocity dispersion has been used in previous studies of the kinematics of dwarf spheroidal galaxies and globular clusters~\citep{2018RPPh...81e6901S}. Here we also include the second term which depends on the tangential velocities in order to extract information from the rotation of the stellar distribution. We choose to assume a gaussian likelihood and bin the velocities in this manner primarily for computational convenience. We find that using an ``unbinned" likelihood and evaluating the CJAM model and thus the likelihood function at the position of each star would be too computationally expensive for our model to converge. 

\par There are two fits to the \gaia\ data performed in this paper. For one, we use both the dispersion and the velocity term in Equation~\ref{eq:vtanlikelihood} to constrain a rotation parameter and the mass distribution. For the second one, we fit to the \gaia\ data again using the $\kappa$ values listed in Table~\ref{tab:surfaceBrightnessMGE} and only fit to the dispersions to constrain the mass distribution. For the HST and RV datasets, we again set $\kappa$ to the values listed in Table~\ref{tab:surfaceBrightnessMGE}~and only fit to the dispersions. By doing so, we have a consistent analysis between all four datasets and also one unique analysis of the \gaia\ dataset in which we explore the effects of fitting to the tangential velocities. 

\par The best fit values of our models are determined by maximizing Equation~\ref{eq:vtanlikelihood} via the MultiNest nested sampling algorithm. We use PyMultiNest~\citep{2016ascl.soft06005B}, the python wrapper of the Bayesian inference tool and nested sampler MultiNest (\citet{2008MNRAS.384..449F},~\citet{2009MNRAS.398.1601F},~\citet{2019OJAp....2E..10F}), to produce posterior probabilities of our free parameters and determine best fit parameters for our models. MultiNest performs nested sampling, in which values are selected from the prior volume and ranked by their likelihoods, $\cal{L}$. The samples with the lowest likelihood, $\cal{L}_{0}$, are thrown out and replaced by a new sample such that $\cal{L}_\text{new} > \cal{L}_0$ and this process is repeated until the entire prior volume has been traversed. Once these calculations are complete, we are left with a list of sampled points that passed the likelihood criteria for each free parameter. These samples can then be used to calculate means and standard deviations for each parameter, as well as various other quantities which depend upon these parameters. For a more thorough description of this process, refer to~\citet{2009MNRAS.398.1601F}. For all calculations which use the MultiNest posterior samples, we use the uniformly weighted samples, given in the~\texttt{post}\_\texttt{equal}\_\texttt{weights.dat} file, an output which is generated upon the completion of the MultiNest routine. \par Table 2 lists the free parameters and prior ranges used for all of our models. For models with no DM, there is no Gaussian fitting performed other than to the surface brightness profile, and so only the mass-to-light ratio and velocity anisotropy are sampled over. 

\par For the velocity anisotropy, the mass-to-light ratio, and the maximum circular velocity, we use uniform linear priors. For $r_{\text{max}}$ we use uniform priors in logarithmic space. For the case of a DM model in which the halo is described by a single Gaussian component, the total mass and Gaussian width are described by uniform priors in logarithmic space. For the case of the $\textit{Gaia}$ dataset, we also set $\kappa$ as a free parameter with a uniform prior range of $[-1,1]$ for all models. Note that in this case, we assume $\kappa$ is the same value for every Gaussian component.  
 \par 
 For all of our models, we keep the inclination angle and distance fixed, choosing to use the inclination angle found by~\citet{2015ApJ...803...29W}, $i = 50^{\circ}$.

We compare our models for each dataset by taking the difference between the log marginal likelihoods as calculated by MultiNest,
\begin{equation}
 \ln(\mathcal{B}) = \ln{(\mathcal{Z}_1)} - \ln{(\mathcal{Z}_0)}
\end{equation}
where $\mathcal{B}$ is the Bayes Factor, traditionally defined as the ratio between the evidences, $\mathcal{Z}_1$ and $\mathcal{Z}_0$ for two models, H$_1$ and H$_0$. In MultiNest, the evidence is defined as the average of the likelihood over the prior volume~\citep{2009MNRAS.398.1601F}, 
\begin{equation}
 \mathcal{Z} = \int \mathcal{L}(\Phi)\pi (\Phi) d^D (\Phi),
\end{equation}
where $\mathcal{L}$ is the likelihood and $\pi$ is the prior, both of which are functions of the free parameters in the model, $\Phi$. Lastly, $D$ is the number of parameters in our model. The simplicity of the model (number of free parameters) is considered in the evidence calculation, and a model with a greater number of free parameters will have a smaller evidence value unless it is significantly favored. We take a Bayes factor greater than $\sim 2$ to be strong evidence for model $H_1$, while a Bayes Factor less than 2 is considered to support the simpler model, $H_0$~\citep{1939thpr.book.....J}.
\begin{table}
	\centering
	\caption{Free parameters and prior ranges for our CJAM+MultiNest models.}
	\label{tab:models_table}
	\begin{tabular}{cc} 
		\hline
		Free Parameter & Prior\\
        \hline
        
        \textbf{NFW} \\
        
		V$_{\text{max}}$ [km/s] &  [5, 50] \\
		Log(r$_{\text{max}}$ [parsecs]) & [$-1$, $2$] \\
		$\Upsilon_{*}$ & [1, 5] \\
		$\beta$ & [-1,1] \\
        
        \textbf{Gaussian} \\
        
        Log(Halo Mass[M$_\odot$]) &  [$3$, $7$] \\
		Log(Gaussian Width [parsecs]) & [$-1$, $2$] \\
		$\Upsilon_{*}$ & [1, 5] \\
		$\beta$ & [-1,1] \\
		
		\textbf{No Dark Mass} \\
		$\Upsilon_{*}$ & [1,5] \\
		$\beta$ & [-1,1] \\
		\hline
	\end{tabular}
\end{table}

\begin{figure*} 
    \includegraphics[width=\textwidth]{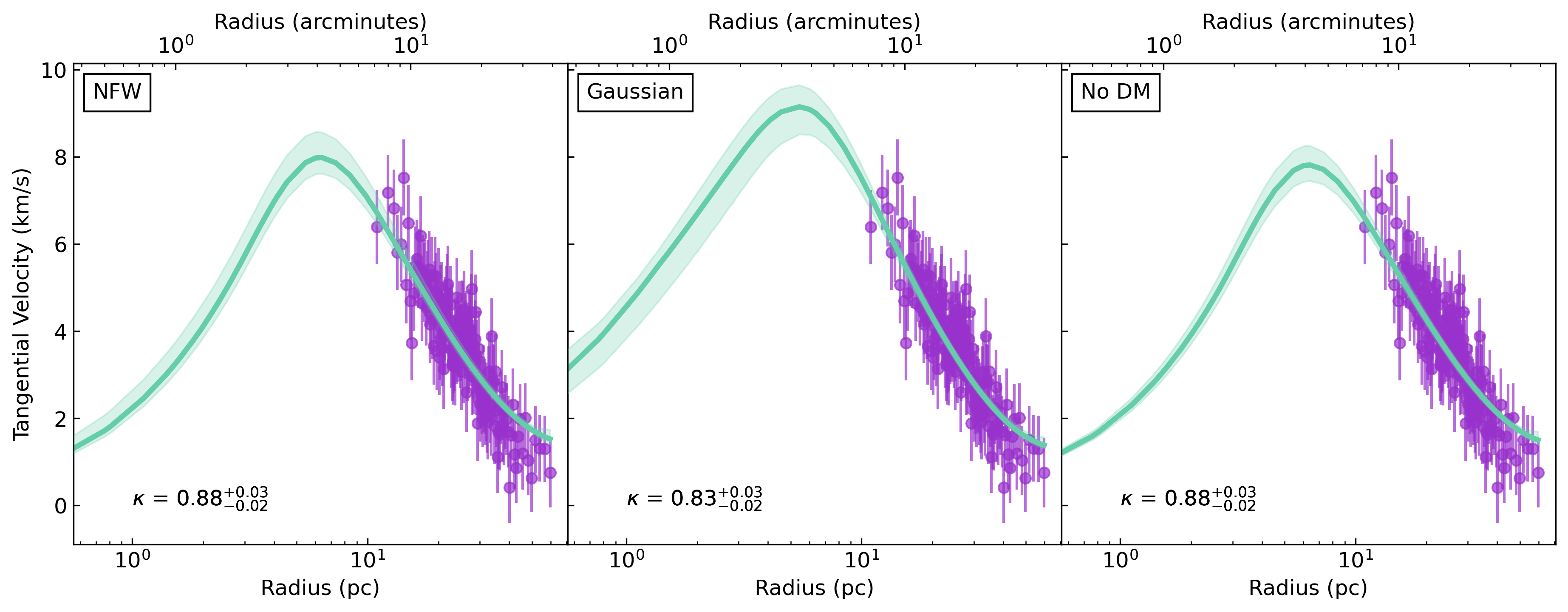}
    \caption{The results of our $\vtan$ + v$_{\text{disp}}$ likelihood analysis for all models (NFW, Gaussian, and No DM) for the \textit{Gaia} dataset with $\kappa$ freed. The purple points show the observed tangential velocities and the green band shows the 90\% containment region of the model values calculated from the MultiNest posterior distributions. The mean values of the posteriors are shown as a green line. The mean $\kappa$ value is also listed for the models in their respective panels.}
    \label{fig:modelVtans}
\end{figure*}
\begin{figure*} 
    \includegraphics[width=\textwidth]{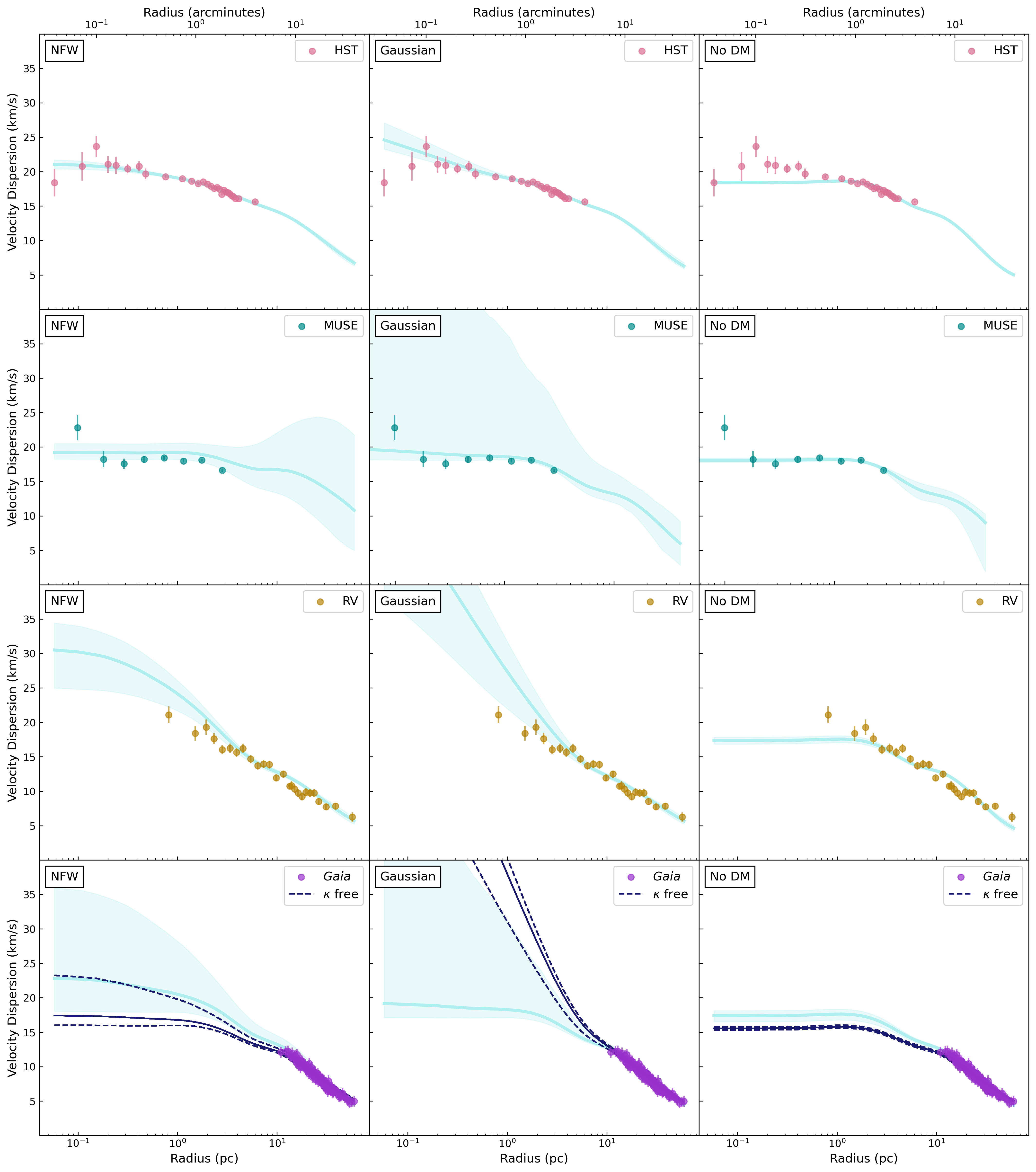}
    \caption{The results of our likelihood analysis for all models and datasets. The colored points show the measured dispersions of each dataset, as shown together in Fig~\ref{fig:observed_dispersions}. HST data is shown as pink points in the top panel, MUSE as cyan points in the second panel, RV as yellow points in the third panel, and \textit{Gaia} as purple points in the bottom panel. The light blue band shows the 90\% containment region of the model velocity dispersion values based off of the MultiNest posterior distributions. The blue line shows the mean of the posterior distributions. The dispersion results from the \gaia\ analysis in which we free $\kappa$ is shown with dark blue lines in the same panel as the $\kappa$ fixed case for comparison. The dashed lines show the 90\% containment region and the solid line shows the mean of the posterior distributions.}
    \label{fig:modelDispersions}
\end{figure*}
\section{Results}
\label{sec:results}
\par We now move on to presenting the results of our analysis. We begin by discussing constraints on the rotation using the tangential velocities and velocity dispersions of EDR3. We then move on to presenting the results for the mass profiles using the velocity dispersions from all four datasets. Finally we use these results to present the implied updated j-factors.  

\label{sec:results}
\subsection{Tangential velocities and rotation parameter}
\label{sec:vtanresults}
We first discuss our analysis of the~\textit{Gaia} data, focusing on fitting to the rotation of~\wcen~with the combination of the tangential velocity and velocity dispersion data. We set $\kappa$ as a free parameter and simultaneously fit to the tangential velocity and dispersion data using equation~\ref{eq:vtanlikelihood}. 
This fitting is performed for the models without DM, for the NFW DM model, and for the Gaussian DM model. For all of these models, we find median values of $\kappa \simeq [0.83 - 0.90] \pm \mathcal{O}(0.01)$.

\par For comparison to these results, we also perform a fit in which our likelihood function contains only the term that depends on the tangential velocity data.  In this case, we find similar values as derived above, $\kappa \simeq 0.90$. This implies that, as expected, $\kappa$ is primarily sensitive to the tangential velocity, and is less sensitive to the remaining dynamical quantities in the jeans equation that determine the mass distribution. 
\par Figure~\ref{fig:modelVtans}  shows the results for the tangential velocity as a function of radius. Though the derived values of $\kappa$ are very similar for all models, the shape of the predicted tangential velocity distribution depends on the chosen model. There are clear differences between our DM models at small radii which are unexplored by the \textit{Gaia} observations, as at these radii~\textit{Gaia} loses sensitivity to proper motions in~\wcen~because of crowding. At these radii, the slopes of the tangential velocity curves differ, and in particular, the no DM model has a significantly smaller peak than models with DM. This implies that larger samples of proper motions within the half-light radius of~\wcen~will be important to constrain the axisymmetric CJAM models that we consider.

\par When simultaneously fitting to $\kappa$ with the tangential velocities, we find that velocity anisotropy is driven to an isotropic value: $\beta \rightarrow 0$. This value of $\beta$ differs from the value derived for the fits to the velocity dispersion with the \textit{Gaia} dataset, which indicates $\beta \sim 0.25$. For fits to the tangential velocity, there is a strong correlation between $\beta$ and $\kappa$, in the sense that increasing $\beta$ results in an increase in $\kappa$. 
There is an additional degeneracy between the rotation parameter, the velocity anisotropy, and the inclination angle of the system, $i$. However, the inclination of our models is restricted by the MGE description of the surface brightness profile, and more specifically by the flattening ratios which are fixed in our models (see equation~\ref{eq:flatteningratio}), and so an in-depth analysis of these relations is beyond the scope of this paper.
Nonetheless, it is clear that degeneracies exist between these three parameters and that a better understanding of them could tell us more about the rotation and anisotropy of a given object. 

\par 
Our result for $\beta$ is consistent with the value derived by~\citet{2013MNRAS.429.1887D}, though these authors determined values of $\kappa$ closer to zero for the majority of the Gaussian components of the MGE. The authors derived $\kappa$ values for each Gaussian component, as shown in our Table~\ref{table:1}, by comparing the rotation derived from their proper motions data from~\citet{2000A&A...360..472V}~to rotation predicted by their axisymmetric Jeans modeling. The authors then adjusted the $\kappa$ values of each component until their model curves matched sufficiently well with the rotation curves interpolated from the data. It is worth mentioning that, in the analysis of~\citep{2013MNRAS.429.1887D}, the velocities at radii $> 19^\prime$ are forced to zero. This differs from our analysis in which the rotation curves extends well beyond this radius. 

\par Rotation has previously been detected in~\wcen~ using spherical Jeans modeling~\citep{2019MNRAS.485.1460S}. Our results show the same counterclockwise rotation for~\wcen~as shown in~\citet{2018MNRAS.481.2125B}~and~\citetalias{2021arXiv210209568V}.
~\citet{2018MNRAS.481.2125B}~uses data from the \textit{Gaia} second data release to calculate proper motion rotation profiles for globular clusters with a global rotation signature greater than 3$\sigma$. 
These authors fit to a rotational component in the tangential direction that is modelled by 
\begin{equation}
    \mu_t = \frac{2V_{{\text{peak}}}}{R_{\text{peak}}} \frac{R}{1 + \left(R/R_{\text{peak}}\right)^2}, 
\end{equation}
where the authors assign a value of $V_{\text{peak}} = 1$ km/s and $R_{\text{peak}} = R_h$. 
From this methodology, the authors find a velocity curve similar to our Figure~\ref{fig:modelVtans}, with the peak of the curve appearing near the same location as ours and with a similar maximum velocity.
\par ~\citetalias{2021arXiv210209568V}~ also look for global rotation in Milky Way globular clusters with \textit{Gaia} EDR3. In this case, the authors use a multivariate Gaussian to describe the joint probability of the proper motions of stars in the cluster. This description is then maximized to determine the best fit velocities. 
Here we do note a small difference in the modeled rotation, in both the peak amplitude (ours is roughly 1-2 km/s larger) and the location of the peak amplitude (ours is located at $\sim4$ arcminutes versus theirs at $\sim7$ arcminutes). 
However, we attribute this difference primarily to the different parameterizations of the rotation curve.
A true one-to-one comparison is difficult as \citetalias{2021arXiv210209568V} does not directly report the rotational PM from the data directly, only their best-fit model.
As a rough proxy, one can use earlier results from \gaia\ DR2 published in \citet{2018MNRAS.481.2125B} where they use a similar parameterization as \citetalias{2021arXiv210209568V}.
They recover a similar rotational amplitude of $\sim6$ km/s but one can see that the data-only values peak above the curve, closer to $\sim7$ km/s, as we find in our data, leaving the parameterization between their work and our work as the main difference and likely driver of the different model peak and peak location.

\subsection{Velocity dispersion profiles}
\par  Figure~\ref{fig:modelDispersions} show the results for the best-fitting velocity dispersions for all four datasets. Here, the light blue band show fits to only the velocity dispersion with the rotation parameter values set to those shown in Table~\ref{tab:surfaceBrightnessMGE}. The dark blue dashed lines in the \gaia\ panels show fits to the data with $\kappa$ freed. 
This figure shows that both the NFW and Gaussian distributions provide good fits over the range of radii probed by the datasets. On the other hand, the velocity dispersions in the no DM case under predict the RV, HST, and and MUSE datasets in the innermost regions. The fits using the Gaussian distribution lead to a sharp rise in dispersions for the HST, RV, and \gaia\ $\kappa$ freed data, while it gives unconstrained results on the \gaia\ $\kappa$ fixed and MUSE datasets.
We note that the results of the fits for the NFW and Gaussian \gaia\ cases, as well as the Gaussian MUSE case, are consistent with the results of the no DM case.

\par Our model with the fewest free parameters is the model with no DM component, so we take this to be our fiducial model and compare the natural logarithm of the Bayes factor for the NFW and Gaussian models relative to this model. For the NFW model, we calculate $\ln(\mathcal{B})_{\text{HST}} = 35.0$ and $\ln(\mathcal{B})_{\text{RV}} = 15.0$. For the Gaussian model, we calculate $\ln(\mathcal{B})_{\text{HST}} = 32.2$ and $\ln(\mathcal{B})_{\text{RV}} = 15.9$. These values suggest a high statistical preference for models which include an extended dark mass. This preference can also be seen in the HST and RV rows of Figure~\ref{fig:modelDispersions}. 
For the MUSE data, we find no statistical preference for the NFW and Gaussian models, with $\ln(\mathcal{B})_{\text{MUSE}} \sim 0$. We note that the NFW and no DM models are unable to fit to the inner-most, high-dispersion data point in the MUSE data. We discuss fitting the HST, RV, and MUSE data with a black hole model below.

\par 
It is interesting to note that the very outer regions of the cluster, which are probed by the $\textit{Gaia}$ data, is the least affected by the presence or non-presence of DM (see the bottom row of Figure~\ref{fig:modelDispersions}), although in the case of \gaia\ $\kappa$ freed we find a slight preference for the Gaussian model with $\ln(\mathcal{B}) = 5.3$. However, the predictions by this model for the dispersions in the inner region of the cluster are unrealistic. Otherwise, we find no preference for an extended dark mass for the \gaia\ data. 
\par For our Gaussian models, we keep $\sigma$ as a free parameter. However, our likelihood analysis is not able to place a lower bound on this parameter, which is set by the input prior. We have examined the limit of small $\sigma$ to model the effect of an IMBH and determine whether this is consistent with the~\wcen~mass distribution. We specifically fix $M_\text{BH} = 10^5$ M$_\odot$ and $\sigma = 1$ arcseconds, and fit to the HST data while freeing the parameters $\Upsilon$ and $\beta$. We compare to the No DM case and find a Bayes factor of $\ln(\mathcal{B}) = 24.5$.
Comparing the evidence for the IMBH model to the model with the extended mass distributions, we find that the evidence for an extended dark mass component is marginally more preferred. We repeat this test again for the RV dataset, this time trying $M_\text{BH} = 10^6$ M$_\odot$ as well as $M_\text{BH} = 10^5$ M$_\odot$ and keeping $\sigma = 1$ arcseconds. Compared to the No DM case for this dataset, the $10^6$ M$_\odot$ IMBH is strongly disfavored. For the $10^5$ M$_\odot$ IMBH, we find a Bayes factor of $\ln(\mathcal{B}) = 11.2$. This again suggests that the extended dark mass is marginally preferred over a model with a $10^5$ M$_\odot$ IMBH. 

Lastly, we test the IMBH scenario again with the MUSE data. As stated previously, we found no preference for the NFW and Gaussian models compared to the no DM model with this dataset. However, the no DM model cannot replicate the inner high-dispersion data point. We compare the Bayes Factors for models in which we add an IMBH with masses of $M_\text{BH} = 10^4$ M$_\odot$ and $M_\text{BH} = 10^5$ M$_\odot$. While we again find no preference for the $10^4$ M$_\odot$ black hole model, the $10^5$ M$_\odot$ black hole model is statistically preferred with $\ln(\mathcal{B}) = 9.0$. While this is not as significant as other models have been found to be in this analysis, it is still clear that the $10^5$ M$_\odot$ is statistically preferred to the no DM case for the MUSE data. In consideration of the shape of the MUSE dispersion profile, it is not surprising that the dispersions are well fit by a high mass point source rather than an extended dark component. We note again the high scatter in dispersions across datasets for the central region of the cluster; see e.g. figure 2 in~\citet{2019MNRAS.482.4713Z}.

\subsection{Mass distribution} 
\par
\begin{figure*} 
    \includegraphics[width=\textwidth]{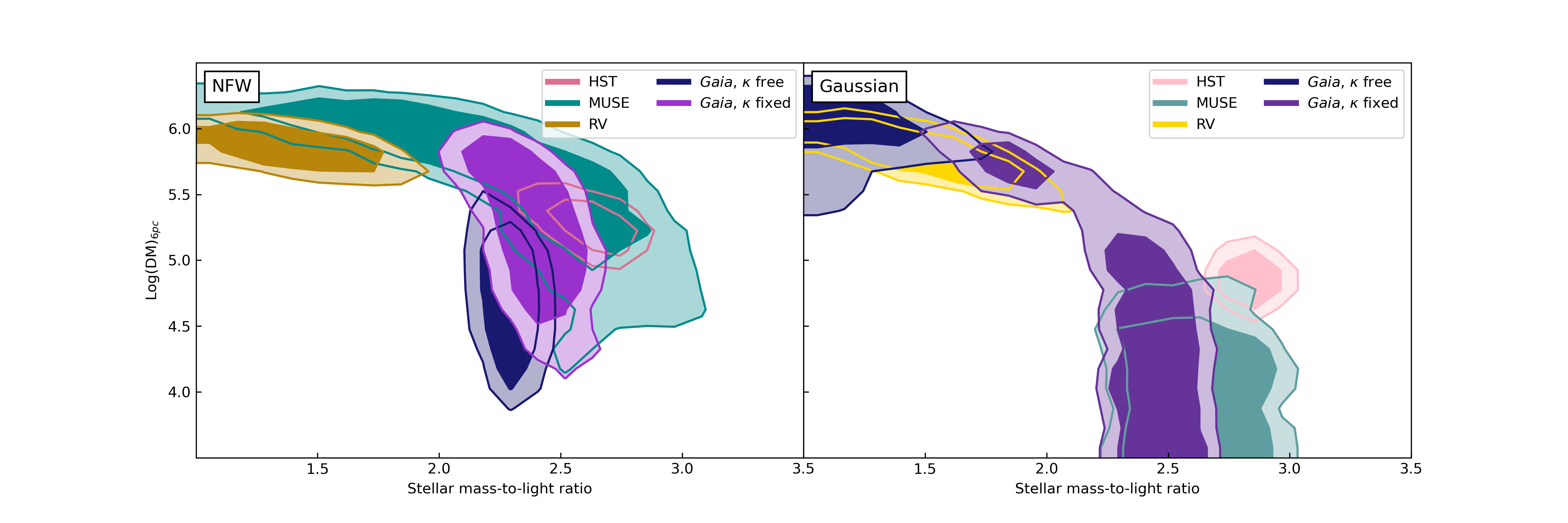}
    \caption{Probability density contours of the stellar mass-to-light ratio versus the dark mass within 6 pc of the center of the cluster for NFW (left) and Gaussian (right) density distributions. The lines denote the 68\% and 90\% containment regions. HST results are shown with different shades of pink, MUSE is shown in shades of cyan, and RV is shown in shades of yellow yellow. The \gaia\ analyses in which we set the $\kappa$ values to the one in Table~\ref{tab:surfaceBrightnessMGE} ($\kappa$ fixed) are shown in shades of purple. The \gaia\ analyses in which we free the rotation parameter ($\kappa$ free) are shown as dark blue in both panels.}
    \label{fig:massContour}
\end{figure*}

 Figure~\ref{fig:massContour} shows the posterior distributions for the mass-to-light ratio $\Upsilon$ versus the mass of the dark component within the half light radius of $6$ pc.  The contours have been calculated by using the MultiNest uniformly weighted posterior samples from our runs and the plotting software Corner~\citep{corner}. 
This figure shows that, for the models that allow a DM component, a non-luminous mass component separate from the stellar potential ranging from $10^4$ to $10^6$ M$_\odot$ is preferred in fitting to a majority of the datasets, except for the cases of the MUSE data and \gaia\ fit with a fixed $\kappa$, which has an unconstrained non-luminous mass. This trend can also be seen in the same cases in Figure~\ref{fig:modelDispersions}, where it is shown that these model + dataset combinations have the largest containment regions. For the NFW case, the MUSE data is also quite unconstrained, which is likely driven by the ill-constrained stellar mass-to-light ratio and velocity anisotropy of this model. 
It is interesting to note that whether or not $\kappa$ is freed in the \gaia\ analysis does not have much an affect on the dark mass for the NFW case, though it significantly changes the dark mass in the Gaussian case. This shift in the dark component's mass depends on the relationship between $\beta$, $\kappa$, and $\Upsilon$. 
\par
\begin{figure*} 
    \includegraphics[width=\textwidth]{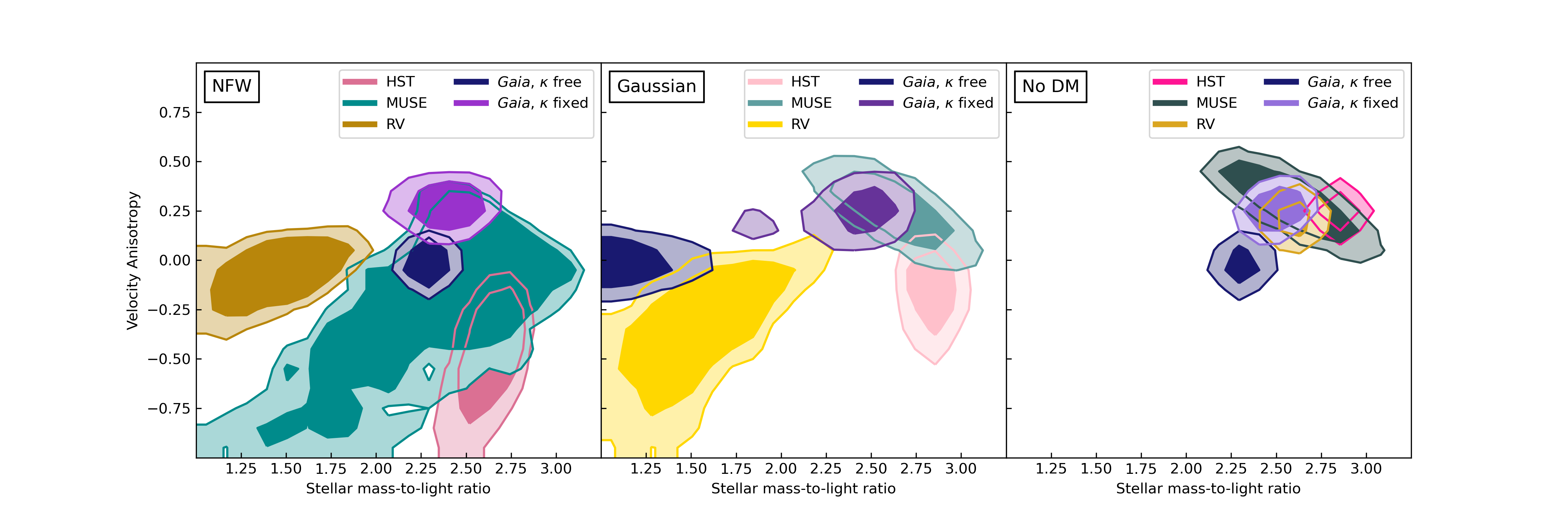}
    \caption{Probability density contours of the stellar mass-to-light ratio versus the velocity anisotropy for a given dataset and model. NFW models are shown on the left panel, Gaussian models in the middle panel, and models with No DM on the right panel. The lines denote the 68\% and 90\% containment regions. HST results are shown with different shades of pink, MUSE is shown in shades of cyan, and RV is shown in shades of yellow yellow. The \gaia\ analyses in which we set the $\kappa$ values to the one in Table~\ref{tab:surfaceBrightnessMGE} ($\kappa$ fixed) are shown in shades of purple. The \gaia\ analyses in which we free the rotation parameter ($\kappa$ free) are shown as dark blue in both panels.}
    \label{fig:MLvsBeta}
\end{figure*}

 As discussed in~\ref{sec:vtanresults}, the value of $\beta$ is driven to zero when $\kappa$ is freed in the fit to the \gaia\ tangential velocity data. This shift in $\beta$ in turn shifts $\Upsilon$, which shifts the mass in the DM component due to the degeneracy between these parameters. The change in the DM mass distribution when fitting with the tangential velocity data as opposed to when it is not included in the fit is most substantial in the Gaussian model. When fitting to the tangential velocity data, the Gaussian model gives a DM mass of $\sim 10^6$ M$_\odot$, as opposed to much lower values of $\lesssim 10^5$ M$_\odot$ when it is not included. In the former case, $\beta \rightarrow 0$ and the DM becomes larger. In contrast, for both the NFW and the no DM model, the DM mass distribution is not significantly affected by fits to the tangential velocity data. We report the values for the velocity anisotropy achieved from our fitting method in Figure~\ref{fig:MLvsBeta}. The $\beta$ and $\Upsilon$ values generally differ for each data set in which a dark component is included in the modeling, spanning over a range of -0.5 to 0.3 for $\beta$ and 1 to $>$ 3 for $\Upsilon$. We note that our modeling is restricted by the assumption of a constant $\beta$ and $\Upsilon$ over all radii. 

\par
\begin{figure*} 
    \centering
    \includegraphics[width=\textwidth]{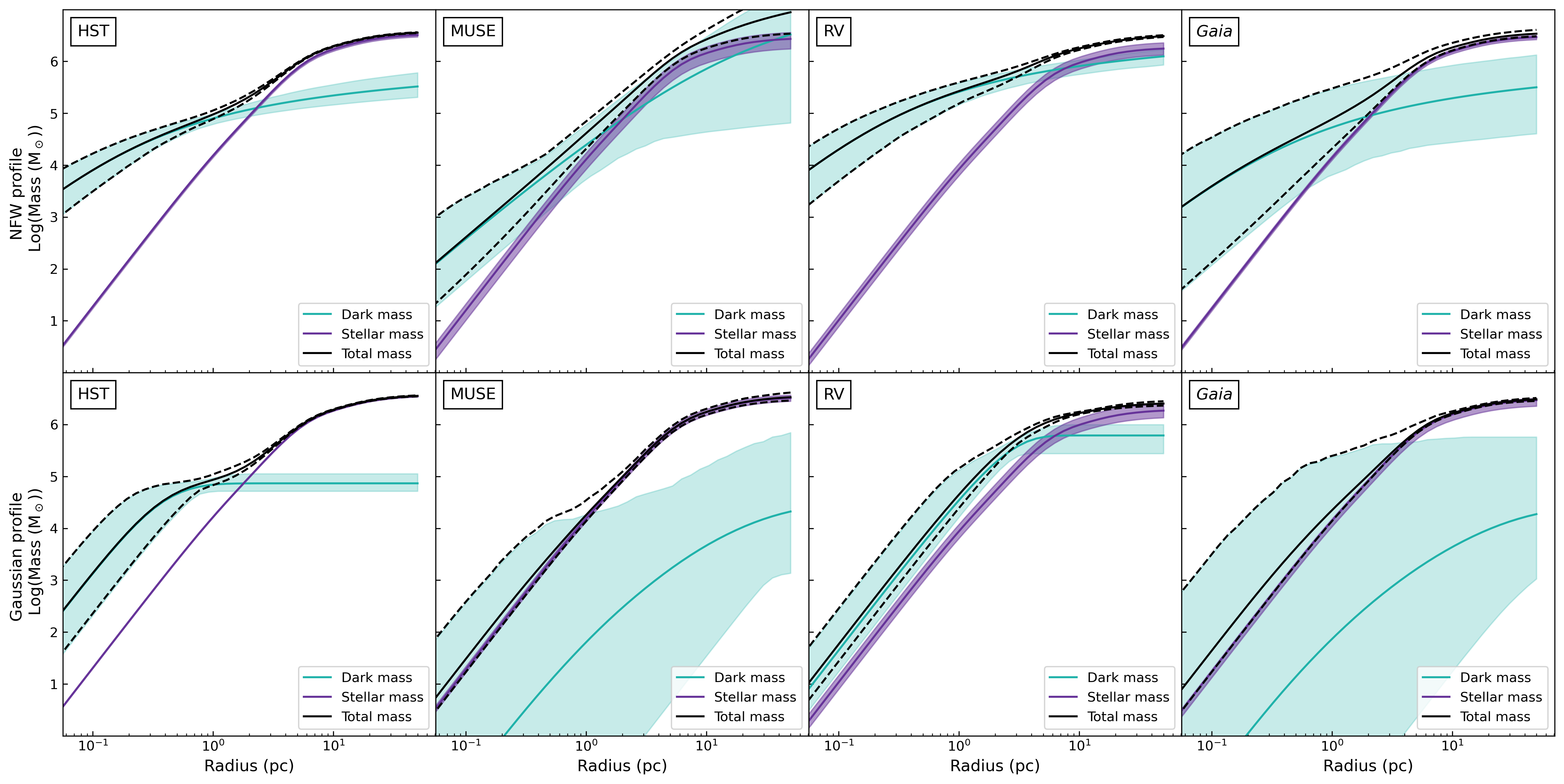}
    \caption{Total mass, dark mass, and spherically averaged stellar mass calculated using the results of our likelihood analysis. The NFW case is shown in the top panel and the Gaussian case is shown in the bottom panel. HST results are shown in the left panel, RV in the middle panel, and \textit{Gaia} in the right panel. The green bands show the 90\% containment region for the dark mass. Likewise, the purple bands show the 90\% containment region for the stellar mass and the dashed black lines show the 90\% containment region for the combined mass. The containment region and mean for the total mass has been calculated by first summing the dark and stellar masses and then calculating the means and percentiles of those distributions.}
    \label{fig:totalMass}
\end{figure*}

 Figure~\ref{fig:totalMass} shows the integrated mass profiles for both the stellar and DM components, for the fits to all datasets with $\kappa$ fixed. For the stellar distributions, we use the spherically averaged stellar mass as defined in Equation~\ref{eq:sphericallyaveragedmass}. For the HST and the RV data, for both of the density profiles, the DM component clearly dominates in the central region within $\sim r_h$. For the MUSE data, the dark mass is poorly constrained compared to the RV and HST results. While the Gaussian dark mass is completely unconstrained, it can be seen that the NFW mass is most poorly constrained at larger radii, which is most likely a result of the radial span of the MUSE data. We also note that the NFW MUSE case shows the least constrained stellar mass-to-light ratio, as can also be seen in Figure~\ref{fig:MLvsBeta}. For the \gaia\ data, in the case of the NFW profile, the DM component dominates the central region, though the uncertainties on it are much larger as compared to the cases of HST and RV. For the Gaussian model, the DM component is the most weakly constrained at all radii. Note that in all cases, the uncertainties in the total stellar plus DM mass profiles are the smallest at near the half-light radius, which is a general result of jeans theory for spherical systems~\citep{Wolf:2009tu}.
\par
\begin{figure*}
    \centering
    \includegraphics[width=0.75\textwidth]{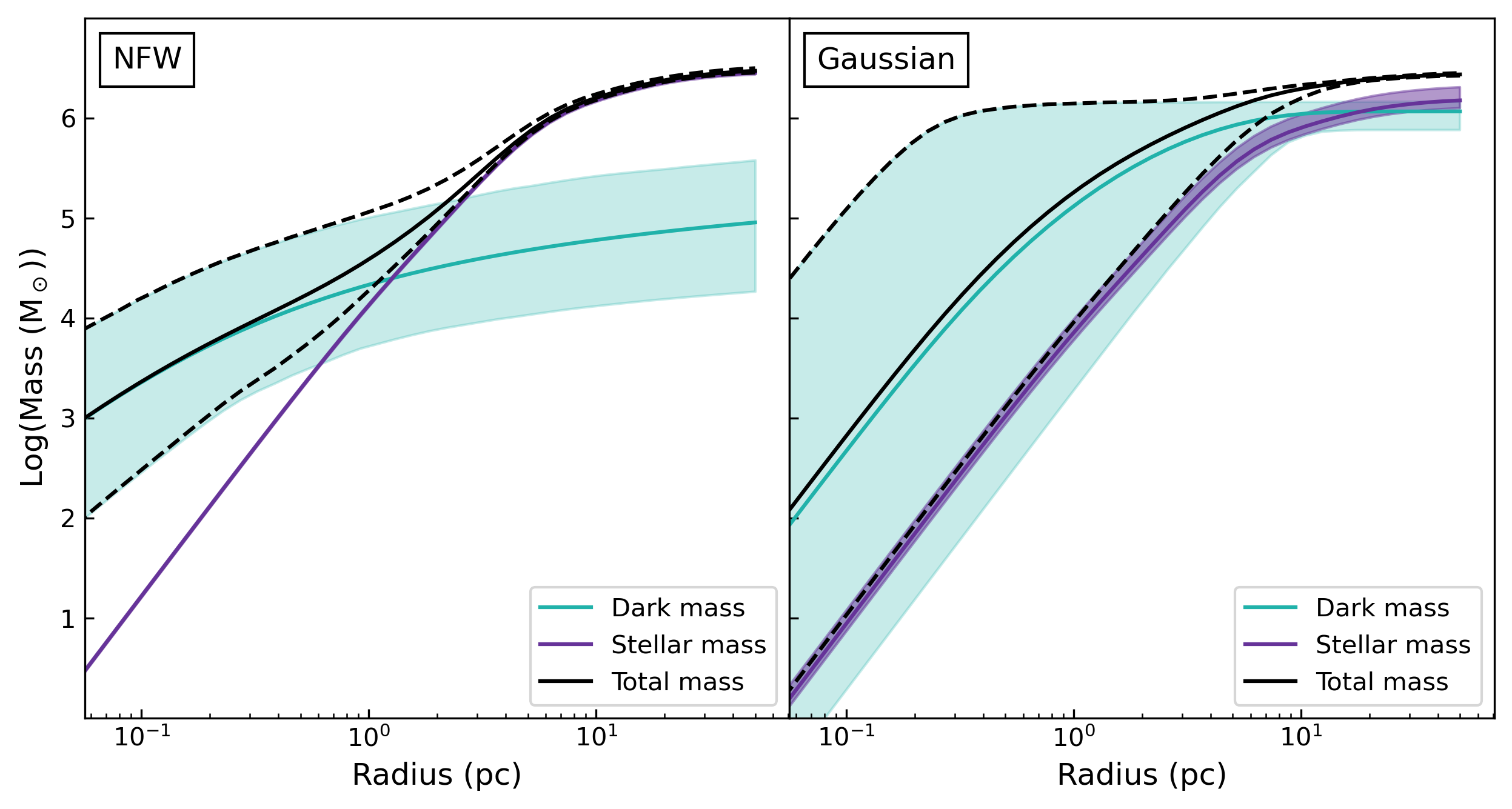}
    \caption{Similar to Figure~\ref{fig:totalMass} except showing the results for \gaia\ with $\kappa$ freed. The NFW case is shown on the left and the Gaussian case is shown on the right. The bands denote 90\% containment regions.}
    \label{fig:totalMass_kappafree}
\end{figure*}

 Figure~\ref{fig:totalMass_kappafree} shows the cumulative mass plots for the \gaia\ dataset with $\kappa$ free. Similar to the results shown in Figures~\ref{fig:modelDispersions}~and~\ref{fig:massContour}, the cumulative mass of the dark component for the NFW profile is consistent with the mass for the \gaia\ $\kappa$ fixed case, whereas the Gaussian profile is more constrained for the $\kappa$ free case. However, the upper limit of the mass range predicted by the \gaia\ $\kappa$ free fits is significantly higher than that predicted by the other datasets and models. 
\par Typical values of \vmax~from our analysis using the NFW profile range from 10 km/s to 40 km/s. The dataset which best constrains the mass of the dark component within the half-light radius is the HST data (see Figure~\ref{fig:totalMass}). The HST data is best fit by a \vmax~of $21.34^{+3.77}_{-2.53}$ km/s. For all datasets the \rmax~is constrained to $ \lesssim 1$ pc. For comparison, for dwarf spheroidals, typical values are~\vmax $\sim 20$ km/s, and \rmax $\sim 1$ kpc~\citep{2014ApJ...783....7C}. 
This shows that the dark mass in~\wcen~is much more centrally-concentrated than the dark matter distribution measured in low mass galaxies. 

\subsection{J-factor and prospects for Indirect Detection}
\par If the dark massive component in~\wcen~is associated with particle dark matter, its relative proximity make it an ideal target to search for particle dark matter annihilation. Even though the integrated dark mass component is less than that associated with a typical dSph by $\sim 2-3$ orders of magnitude, it is over 4 times closer to the Earth than the nearest such dSph that has been used to set limits on dark matter annihilation~\citep{Aleksic:2011jx,Ackermann:2015zua}. 
A high J-factor can be achieved by a system that is the appropriate combination of high density and close proximity. The J-factor is then used to determine the flux due to the creation of neutral particles from DM annihilation, assuming the dark matter annihilates into Standard Model particles. 
\par We determine the J-factor in a standard manner from the results of our MultiNest analysis above~\citep{Strigari:2018bcn,Pace:2018tin}. For the dark mass distributions that we calculate above, the angular extent of the dark component is always $\lesssim 1^\circ$, implying that the spatial distribution of the gamma-ray emission would be that of a point source. In this case, assuming a spherical NFW density distribution, the J-factor may be approximated by~\citep{2016PhRvD..93j3512E} 
\begin{equation}
    J \simeq \frac{4 \pi \rho_s^2 r_s^3}{3 D^2} 
\end{equation}
\par Using the results for the dark mass density distributions derived above, we derive the J-factors for the three datasets that show some statistical preference for an extended dark component.
Under the assumption of an NFW profile, for the HST, RV, and \gaia\ data, the integrated median J-factors are $2.6 \times 10^{22}$, $2.3 \times 10^{24}$, and $1.3 \times 10^{21}$ GeV$^2$ cm$^{-5}$, respectively. The wide range of possible median J-factors reflect the range of possible mass profiles as shown in Figure~\ref{fig:massContour}. Even though there is a wide range of possible values, the J-factor we calculate is larger than that of any of the known dSphs~\citep{Pace:2018tin}. A full determination of the cross sections bounds (or possible signal that may be extracted from~\wcen~) involves implementing this J-factor with its associated uncertainties into gamma-ray data, and accounting for the possible gamma-ray emission from MSPs, which is beyond the scope of this paper. 
\section{Discussion \& Conclusions}
\label{sec:discussion}
In this paper, we have analyzed the dynamics of the~\wcen~globular cluster using the most recent measurements of the internal stellar line-of-sight velocities and proper motions. 
We consider two models for the distribution of the non-luminous dynamical component: an NFW profile and a Gaussian profile. 
We find that for all datasets and for all density profiles, the dynamics of~\wcen~provide evidence for a centrally-concentrated distribution of matter that is distinct from the luminous component, with a mass of $10^4 - 10^6$ M$_\odot$.

\par Though our analysis has been performed for an axisymmetric model, the results for the integrated mass of the non-luminous component is in good agreement with previous spherically-symmetric models~\citep{Brown:2019whs}. This is in spite of the fact that, in the spherical limit, CJAM is restricted to isotropic stellar orbits, making our analysis in the spherical limit a somewhat more restrictive model than the one used in~\citet{Brown:2019whs}. An additional possible restriction of our model is the assumption of a constant mass-to-light ratio, constant velocity anisotropy, and the fact that $\kappa$ is a constant value across all Gaussian components. Though~\citet{Brown:2019whs} find that a variable mass-to-light ratio does not affect the conclusions for spherical models, the addition of it would provide more felixibility for our models. 

This is because there is a clear degeneracy between the dark mass and stellar mass-to-light ratio, as can be seen in Figure~\ref{fig:massContour}, which could be better explored with a variable stellar-mass-to-light ratio. For our constant mass-to-light ratio model with a potential described only by the stellar mass, we find a stellar mass-to-light ratio consistent with the results from~~\citet{2013MNRAS.436.2598W}.

\par From our analysis of Gaia EDR3 tangential velocity data, we provide a new estimate of the rotation of~\wcen. For our assumed model, we find a rotational component extending out to larger radii than was found in previous studies that used the same formalism as presented in this work, such as by~\citet{2013MNRAS.429.1887D}. However we do note that multiple previous \gaia\ studies have found rotation out to similarly large radii \citep[e.g.,][]{2018MNRAS.481.2125B, Jindal2019, 2021arXiv210209568V}. But here our focus is on the inclusion of and the constraining of rotation using the CJAM formalism. This rotation is independent of the assumed mass model, and weakly-dependent on the existence and the nature of the centrally-concentrated dark mass component. The different mass models predict differences in the rotation curve at radii smaller than the EDR3 data is sensitive to. Improved astrometric measurements of tangential velocities in the central region of the system may be able to further distinguish between these models.

\par If the dark mass component is associated with particle dark matter, we calculate the J-factor from particle dark matter annihilation, and show that~\wcen~is likely the most promising target for dark matter constraints. Though the existing gamma-ray flux from~\wcen~may be explained by MSPs, a more detailed study of the MSP contribution could determine whether there is room for a dark matter component, or how the bounds on the cross section compare to those from dSphs. 

\par The analysis in this paper has assumed that the stellar and dark mass distributions are in dynamical equilibrium, so that the mass distribution is faithfully determined by applying equilibrium jeans models to the stellar kinematics and photometry. We believe this is a plausible assumption for the stellar distribution within a limiting radius of $\sim 60$ pc, since the surface brightness is characteristic of globular clusters and dwarf spheroidals, and the velocity dispersions are smooth within this limiting radius. However, it is possible that non-equilibrium features are visible beyond the limiting radius, as high probability member stars associated with~\wcen~have been identified in this region~\citep{2021MNRAS.tmp.2080K}. More accurately mapping out the velocity dispersions at and beyond the limiting radius may be useful in mapping to the transition from  equilibrium to non-equilibrium dynamics.  
\par While a centrally-concentrated core of dark matter is consistent with~\wcen's~kinematics, another possibility is that the dark massive component is a population of stellar remnants within the half-light radius of~\wcen. Following a similar calculation performed in~\citet{axionDMpaper}, we use the results of~\citet{Kremer2020_GCpaper} to estimate the mass due to stellar remnants given a luminous mass of~\wcen~of $\sim 2 \times 10^6 M_\odot$.~\citet{Kremer2020_GCpaper}~uses stellar-dynamical Monte Carlo models to estimate the number of stellar remnants in a Milky Way GC given a number of initial parameters such as the cluster's metallicity, virial radius, number of stars, and distance from the galactic center.

For a GC with the properties of~\wcen~, the final number of white dwarf remnants is $2 \times 10^5$, of neutron star remnants is $2 \times 10^3$, and of black holes is $2 \times 10^2$. Scaling to the mean masses of these remnants, we estimate a total mass in stellar remnants possibly as large as $M_{\text{stellar remnants}} \sim 5 \times 10^5 M_\odot$. This mass is consistent with some of our implied mass range, though is not able to accommodate the large mass regime $\lesssim 10^6$ M$_\odot$. Moreover, this estimate may be consistent with that of stellar remnants in the form of stellar mass black holes~\citep{2019MNRAS.488.5340B}. This highlights the need for even further improved kinematic data in determining the nature of the dark mass component in~\wcen. If the extended dark mass that our study has found is what is left of \wcen's dark matter halo, then \wcen~most closely resembles an ultra-compact dwarf galaxy, such as the dwarf galaxy M32. This is more clearly shown in~\citet{Brown:2019whs}. The resemblance to M32, thought to be the tidally stripped core of a more massive galaxy~\citep{Dsouza2018}, further supports the hypothesis that \wcen~is the remnant core of a dwarf galaxy.

 \section*{Data availability}
 The data underlying this article will be shared on reasonable request to the corresponding author.
\section*{Acknowledgements}
We thank Peter Ferguson, Jennifer Marshall, and Jonelle Walsh for discussions. We thank Laura Watkins for helpful correspondence and making CJAM publicly available. AE and LES acknowledge support from DOE Grant de-sc0010813. This work was supported by a Development Fellowship from the Texas A$\&$M University System National Laboratories Office.

This work has made use of data from the European Space Agency (ESA) mission
{\it Gaia} (\url{https://www.cosmos.esa.int/gaia}), processed by the {\it Gaia}
Data Processing and Analysis Consortium (DPAC,
\url{https://www.cosmos.esa.int/web/gaia/dpac/consortium}). Funding for the DPAC
has been provided by national institutions, in particular the institutions
participating in the {\it Gaia} Multilateral Agreement.
This work was completed by the use of the \textsc{Python} programming language as well as the following software packages: \textsc{astropy} ~\citep{astropy:2018}, \textsc{pandas}~\citep{reback2020pandas}, \textsc{numpy}~\citep{         harris2020array}, \textsc{scipy}~\citep{2020SciPy-NMeth}, \textsc{matplotlib}~\citep{Hunter:2007}, and \textsc{spyder}~\citep{raybaut2009spyder}.



\bibliographystyle{mnras}
\bibliography{example} 








\bsp	
\label{lastpage}
\end{document}